\documentclass[prd,showkeys,preprintnumbers,floatfix,superscriptaddress,
nofootinbib]{revtex4}
\usepackage{float}
\usepackage{nicefrac}
\usepackage{slashed}
\usepackage{mathtools}
\usepackage{amsfonts} % AMS
\usepackage{amssymb} % AMS
\usepackage{amsmath} % AMS
\usepackage{esint}
\usepackage{graphicx} % Include figure files
\usepackage{subfigure} % Include figure files
\usepackage{array} % array
\usepackage{dcolumn} % Align table columns on decimal point
\usepackage{bm} % bold math
\usepackage{latexsym} % latex symbols
\usepackage{longtable} % long tables
\usepackage{hyperref} % hypertext links 
\usepackage{verbatim}
\usepackage{epsfig}
\usepackage[usenames,dvipsnames,svgnames,table]{xcolor}
\newcommand{\M}[0]{ m }

\begin{document}
 \preprint{\vbox{\hbox{JLAB-THY-17-2434} }}
\title{Role of the Euclidean signature in lattice calculations 
\\ of quasidistributions and other nonlocal matrix elements}

%%%%%%%%%%
\author{Ra\'ul A.~Brice\~no}
\email[e-mail: ]{rbriceno@jlab.org}
\affiliation{
Thomas Jefferson National Accelerator Facility, 12000 Jefferson Avenue, Newport News, Virgina 23606, USA\\
}
%%%%%%%%%%

%%%%%%%%%%

\author{Maxwell T.~Hansen}
\email[e-mail: ]{hansen@kph.uni-mainz.de}
\affiliation{
Institut f\"ur Kernphysik and Helmholtz Institute Mainz, Johannes Gutenberg-Universit\"at Mainz,
55099 Mainz, Germany\\
}
%%%%%%%%%%

%%%%%%%%%%
\author{Christopher J.~Monahan}
\email[e-mail: ]{chris.monahan@rutgers.edu}
\affiliation{
New High Energy Theory Center and Department of Physics and Astronomy,
Rutgers, The State University of New Jersey, Piscataway, New Jersey 08854, USA\\
}
%%%%%%%%%%

\date{\today}

\begin{abstract}
Lattice quantum chromodynamics (QCD) provides the only known systematic, nonperturbative method for first-principles calculations of nucleon structure. However, for quantities such as light-front parton distribution functions (PDFs) and generalized parton distributions (GPDs), the restriction to Euclidean time prevents direct calculation of the desired observable. Recently, progress has been made in relating these quantities to matrix elements of spatially nonlocal, zero-time operators, referred to as quasidistributions. Still, even for these time-independent matrix elements, potential subtleties have been identified in the role of the Euclidean signature. In this work, we investigate the analytic behavior of spatially nonlocal correlation functions and demonstrate that the matrix elements obtained from Euclidean lattice QCD are identical to those obtained using the Lehmann-Symanzik-Zimmermann reduction formula in Minkowski space. After arguing the equivalence on general grounds, we also show that it holds in a perturbative calculation, where special care is needed to identify the lattice prediction. Finally we present a proof of the uniqueness of the matrix elements obtained from Minkowski and Euclidean correlation functions to all order in perturbation theory. 
 \end{abstract}


\nopagebreak
\maketitle

\section{Introduction}

Quantum chromodynamics (QCD), the gauge theory of the strong force, is nonperturbative at the hadronic scales relevant for understanding the structure of protons and neutrons. In principle, this obstacle can be navigated using lattice QCD, in which QCD is formulated on a finite and discretized Euclidean spacetime. The correlation functions of the theory are then represented by discretized path integrals that can be estimated stochastically with large-scale numerical calculations.  

Nucleon structure, however, poses a central challenge to this approach. The paradigm examples are parton distribution functions (PDFs), which characterize the longitudinal momentum structure of the nucleon. PDFs are defined as matrix elements of operators extended in the lightlike direction and are consequently inaccessible in Euclidean spacetime, where the light cone is a single point. Traditionally, lattice calculations have attempted to overcome this problem by determining the Mellin moments of PDFs~\cite{Martinelli:1987zd,Gockeler:2000ja,Dolgov:2002zm,Gockeler:2005vw,Hagler:2007xi}, which can be related to matrix elements of local twist-two operators. Unfortunately, this procedure is currently limited to the first few moments of PDFs by power-divergent mixing induced by the reduced symmetry of lattice QCD \cite{Detmold:2001dv,Detmold:2003rq}.

A new route to the direct determination of PDFs from lattice QCD, recently proposed in Ref.~\cite{Ji:2013dva}, is to instead extract so-called quasi-PDFs or quasidistributions. In this approach one considers the matrix element of an operator extended in a spacelike direction between two nucleon states evaluated at finite momentum. This quasi-distribution can be then related to the light-front PDF through a perturbative matching condition~\cite{Xiong:2013bka,Ji:2015qla}, with the effects of the finite nucleon momentum incorporated through an effective theory, LaMET \cite{Ji:2014gla}, or to transverse momentum distributions \cite{Radyushkin:2016hsy,Radyushkin:2017gjd}. This has inspired exploratory calculations of quark quasidistributions~\cite{Alexandrou:2015rja,Chen:2016utp,Zhang:2017bzy}. These results incorporate only a single lattice spacing, and issues of renormalization on the lattice are yet to be fully resolved~\cite{Xiong:2013bka,Ji:2015qla,Ishikawa:2016znu,Chen:2016fxx,Monahan:2016bvm}.

At the heart of this approach, and implicit in the calculations of Refs.~\cite{Xiong:2013bka,Ji:2015qla}, is the intuition that the quasi-PDFs extracted in lattice calculations are equal, up to discretization and finite-volume effects, to those defined using Minkowski signature correlators. In particular, the matching strategy of Refs.~\cite{Xiong:2013bka,Ji:2015qla} assumed that the collinear divergences of the quasi-PDF that appear at one loop in perturbation theory are independent of whether one uses Euclidean-signature or Minkowski-signature correlation functions. The observation that the time-independent matrix element relevant to quasi-PDFs carries no knowledge of the signature of the correlator from which it is determined has not been examined with direct QCD calculations until recently. 
 
Evidence in support of the assumption appeared in \cite{Monahan:2016bvm}, which analyzed the relationship of the closely-related smeared quasidistributions (these differ from quasidistributions only in their ultraviolet behavior) and light-front PDFs via Mellin moments. A more direct perturbative investigation appeared in Ref.~\cite{Carlson:2017gpk}, where it was shown that there are kinematic regions in which the analytic continuation of Feynman diagram loop integrals requires care. From this insight, the authors of Ref.~\cite{Carlson:2017gpk} suggest that the claim that infrared (collinear) divergences of the quasi-distribution in Euclidean spacetime coincide with those of the light front PDF needs careful examination and is possibly misplaced.\footnote{Differing infrared behavior in Euclidean and Minkowski spacetime would cast doubt on the viability of the entire program, although it is worth noting that such infrared behavior is not required in the approach taken in \cite{Ma:2014jla}.}

In this work, we address these issues by analyzing the properties of the Euclidean-time dependent correlation functions that are calculated in lattice QCD. We present a general argument that, for any single-particle matrix element of a current that is local in time (possibly nonlocal in space), the exact same quantity is determined from the long-time behavior of a Euclidean correlator as from the Lehmann-Symanzik-Zimmermann (LSZ) reduction procedure on a Minkowski-space three-point function. As we demonstrate in Sec.~\ref{sec:euclidean}, although the correlation functions carry information on the signature of the spacetime, the matrix elements of time-local operators do not, and therefore these should not be assigned the label ``Euclidean'' or ``Minkowski.'' In Sec.~\ref{ssec:pertth}, we consider the implications of our discussion for the perturbative example posed by Ref.~\cite{Carlson:2017gpk} and resolve the apparent tension between the Euclidean and Minkowski calculations. Finally, in Sec.~\ref{ssec:all_orders} we give a proof, to all orders in perturbation theory, that the matrix element appearing in a Euclidean correlator coincides with the Minkowski analogue. 
 
\begin{figure}[t]
\centering
\includegraphics[width =.8\textwidth]{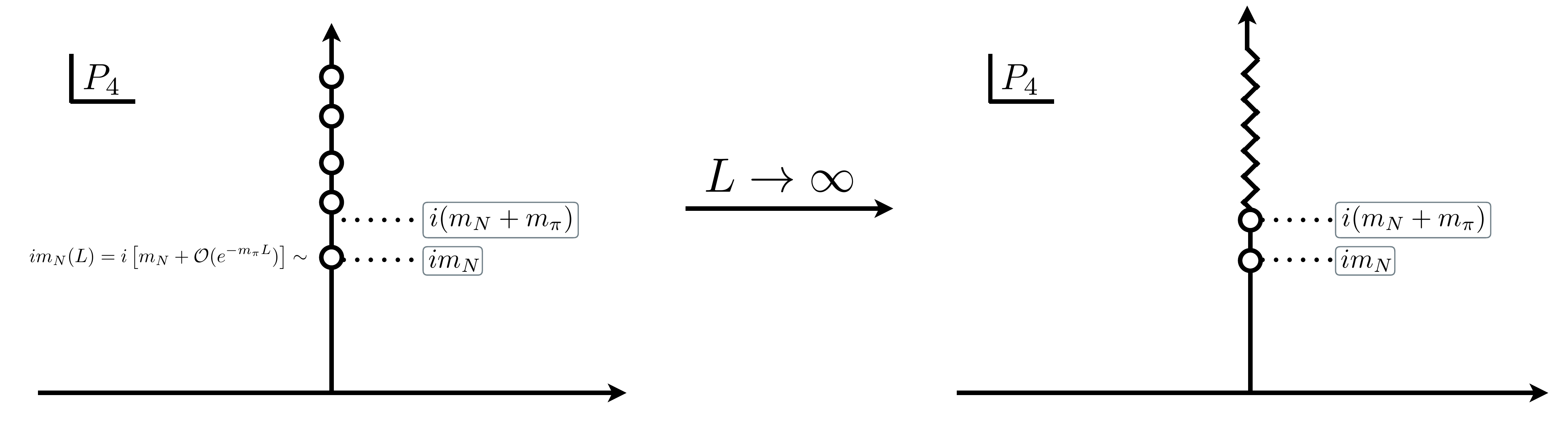}
\caption{The analytic structure for the finite- and infinite-volume correlation functions, left- and right-hand diagrams, respectively, carrying the quantum numbers of a nucleon at rest in a Euclidean spacetime. In both cases we show the first threshold, where a nucleon-pion pair with nucleon quantum numbers can go on shell. In the infinite-volume limit this introduces a cut, as is shown in the right-hand diagram.}\label{fig:corr} 
\end{figure}

\section{General relations for matrix elements and correlation functions\label{sec:euclidean}}
 
Numerical lattice QCD calculations are necessarily restricted to a nonzero lattice spacing, to a finite-volume spacetime and to Euclidean signature correlation functions.\footnote{Many calculations are also performed at unphysical quark masses, but this is not a fundamental requirement and physical point calculations do now exist.} For a given target observable, it is important to understand the effects of these three restrictions.  

The nonzero lattice spacing, denoted $a$, provides a numerically tractable nonperturbative regularization of QCD. One studies the effect of discretization, i.e.~of $a \neq 0$, by calculating the same observable at many different lattice spacings, with the bare parameters tuned so that observables agree up to $\mathcal O(a^n)$ corrections, where usually $n=1$~or~2. Extrapolating $a \to 0$, \textit{i.e.}~taking the continuum limit, then removes the unwanted residual discretization effects. For the purposes of the formal analysis in this work, we will assume that the appropriate continuum limit has been taken. We comment, however, that this procedure is not fully understood for quasidistributions and has been the focus of recent work \cite{Ishikawa:2016znu,Chen:2016fxx,Monahan:2016bvm}.

Similarly, in the analysis presented here we do not include the effects of the finite-volume spacetime, as these are irrelevant for the issues that we aim to understand. The effect of finite Euclidean time is that lattice calculations are in fact thermal averages, but in this work we assume the time extent is taken large enough that nonzero-temperature effects are negligible. This leaves a finite spatial volume, assumed to have periodicity $L$ in each direction. 

An important consequence of finite $L$, shown in Fig.~\ref{fig:corr}, is that the spectrum is discrete, meaning that correlation functions can be written as discrete sums over exponentials decaying in Euclidean time. The decay rates then give the finite-volume energies and the coefficients can be used to extract finite-volume matrix elements. These energies and matrix elements carry $L$ dependence, and in the case of multiparticle states a robust theoretical treatment is required to extract physical observables from the finite-volume quantities~\cite{Luscher:1985dn,Briceno:2014uqa,Briceno:2017tce}. However, in this work we are only interested in single-particle energies and matrix elements. These are known to differ from their infinite-volume counterparts by corrections of the form $e^{- m_\pi L}$~\cite{Luscher:1985dn}, which we take to be negligible.

In Secs.~\ref{ssec:pertth} and \ref{ssec:all_orders} we are concerned with the analytic structure of correlation functions in the complex plane, so it is useful to also directly compare the finite- and infinite-volume versions of these objects. As is shown in Fig.~\ref{fig:corr}, the analytic structures differ significantly, with the multiparticle poles of the finite-volume correlator replaced by a cut in infinite volume. However, the single-particle state manifests as a pole in both cases, and therefore the differing structure elsewhere is irrelevant for our purposes.  In summary, the discussion of Euclidean time and its effect on single-particle matrix elements can be equally well performed in finite and infinite volume. We find it more convenient to work in infinite volume, primarily to simplify notation and to more directly connect to previous studies of parton distributions on the lattice.

Finally, the role of Euclidean time differs crucially from that of the nonzero lattice spacing and the finite volume, in that no parameter can be directly tuned to recover Minkowski correlators from their Euclidean counterparts.\footnote{In principle analytic continuation defines such a parameter, but this cannot be used given only numerical lattice QCD data with finite uncertainties.} Thus, one must either identify quantities that are unaffected by the distinction between Minkowski and Euclidean or else derive relations for converting between the two that can be realistically applied to numerical lattice data. Since no achievable limit connects correlators of different signature, generally one should not expect a situation in which they differ but are numerically close. 

In this work we are concerned with a class of matrix elements that can be accessed from both Minkowksi and Euclidean correlators and, by the nature of their definitions, will be seen to not meaningfully carry a label of Minkowski or Euclidean. To construct these quantities, we begin by introducing the Fock space of a  generic quantum field-theory with only massive degrees of freedom. We restrict attention to single-particle states and denote these by $\vert \textbf P, \mathcal Q \rangle$ where $ \textbf P$ indicates the three-momentum of the state and $\mathcal Q$ all other quantum numbers. We require these single-particle states to have the standard, relativistic normalization
\begin{equation}
\langle \textbf P', \mathcal Q' \vert \textbf P, \mathcal Q \rangle = 2 \omega_{\textbf P} (2 \pi)^3 \delta^3(\textbf P' - \textbf P) \delta_{\mathcal Q' \mathcal Q} \,,
\end{equation} 
where $\omega_{\textbf P} = \sqrt{\textbf P^2 + \M_{\mathcal Q}^2}$ and $\M_{\mathcal Q}$ is the physical pole-mass of the stable particle. We stress that these states are unambiguously defined as eigenstates of the Hamiltonian, $\hat{H}$, with the specified quantum numbers and normalization. In particular, it is not meaningful to attach a label of Minkowski or Euclidean to these states. 

Reference~\cite{Ji:2013dva} proposed that one can numerically evaluate the quark distribution functions of a stable hadron from lattice QCD correlation functions with an insertion of the operator
\begin{equation}
\int \frac{d\xi_z}{4\pi} e^{ix\xi_zP_z}\overline{\psi}(\xi)\gamma_z W[\xi;0]{\psi}(0),
\label{eq:qPDF_op}
\end{equation}
where $\psi$ is a quark field and $W[\xi,0]$ is the Wilson line joining the origin and the spatial point $\xi^\mu=(0,0,0,\xi_z)$. This is a conceptual departure from the operator that most commonly is used for parton distributions
\begin{equation}
\int \frac{d\widetilde\xi_-}{4\pi} e^{ix\widetilde\xi_-P_+}\overline{\psi}(\widetilde\xi)\gamma_+ W[\widetilde\xi;0]{\psi}(0),
\label{eq:PDF_op}
\end{equation}
where $\widetilde{\xi}=(0,\widetilde\xi_-,0)$ is a four-vector in the standard light-front coordinates, $x = (x_+,x_-,x_\perp)\equiv ((z+t)/2,(z-t)/2,x_1,x_2)$. The reason for introducing the operator in Eq.~\eqref{eq:qPDF_op} is that, unlike Eq.~\eqref{eq:PDF_op}, it carries no information about the signature of the spacetime.

In general, one can think of a generic composite operator $\mathcal O(0, \{\pmb \xi \})$ that is localized at zero time, but may be delocalized in space and depend on $n$ displacement vectors, $\{ \pmb \xi \}=\{\pmb \xi_1, \cdots, \pmb \xi_n\}$. The quantity shown in Eq.~\eqref{eq:qPDF_op} is a Fourier transform of an operator of this form. This class of displaced operators is relevant for the study of both parton distribution functions and generalized parton distributions \cite{Ji:2015qla}, via matrix elements of the form
\begin{equation}
\mathcal{M}_{\mathcal Q' \mathcal Q}^{\mathcal O} (\{ \pmb \xi\}, \textbf P', \textbf P) \equiv \langle \textbf P', \mathcal Q'  \vert   \mathcal O(0, \{\pmb \xi \})  \vert \textbf P, \mathcal Q \rangle  \,.
\label{eq:mat_elem}
\end{equation}

Like the energy eigenstates, the zero time operator $\mathcal O(0, \{\pmb \xi \})$ should not be assigned a label of Minkowski or Euclidean. It is localized to the origin of the complex time plane, where the real and imaginary axes cross, and its definition is unaffected by the conventions used in calculating correlation functions. The spacetime signature of operators arises only when these are time-evolved,
\begin{align}
\mathcal O_{M}(t)&\equiv e^{i\hat{H}t} \,\mathcal O(0) \,e^{-i\hat{H}t},\\
\mathcal O_{E}(\tau)&\equiv e^{\hat{H}\tau} \,\mathcal O(0) \, e^{-\hat{H}\tau},
\label{eq:Heis_Euc}
\end{align}
to define Heisenberg-picture operators in Minkowski and Euclidean spacetime.

In summary, neither the energy eigenstate, nor the zero-time operator carry any characteristic that defines them as Minkowski or Euclidean. We thus deduce that the matrix element, $\mathcal{M}_{\mathcal Q' \mathcal Q}^{\mathcal O}$, itself does not carry such a label. In the following subsection we discuss how this signature-agnostic quantity can be calculated using both Euclidean and Minkowski correlation functions. Then, in Sec.~\ref{sec:displaced}, we highlight how displacing the current in time induces a meaningful distinction between Minkowski and Euclidean matrix elements.

\subsection{Calculating matrix elements using lattice QCD}

In a numerical lattice QCD calculation, one stochastically estimates Euclidean correlators using importance sampling techniques. In particular it is possible to calculate two-point functions of the form
 \begin{equation}
C_{ \mathcal Q}(\tau' - \tau, \textbf P) \equiv \langle  \widetilde{N}_{\mathcal Q}(\tau', \textbf P)      N^\dagger_{\mathcal Q}(\tau, \textbf 0) \rangle \,,
\label{eq:two_point}
\end{equation}
where $N^\dagger_{\mathcal Q}(\tau, \textbf 0)$ creates states with quantum numbers $\mathcal Q$ and is localized at the origin in space and at Euclidean time $\tau$. We have also introduced
 \begin{equation}
 \widetilde N_{\mathcal Q'}(\tau', \textbf P) \equiv \int d^3 \textbf x \,   e^{- i \textbf P \cdot \textbf x}   \,   N_{\mathcal Q'}(\tau', \textbf x) \,. 
 \end{equation}

Inserting a complete set of states
\begin{align}
1=\sum_{\mathcal Q} \int \frac{d^3 \textbf P}{  (2\pi)^3 } 
\frac{\vert \textbf P, \mathcal Q \rangle \langle \textbf P', \mathcal Q \vert}{2\omega_{\textbf P}}
+\cdots,
\end{align}
where the ellipsis denotes the contribution from multiparticle states, and also using the Euclidean time translation operators defined in Eq.~\eqref{eq:Heis_Euc}, we find
\begin{equation}
C _{ \mathcal Q}(\tau' - \tau, \textbf P) = 
|  Z_{\mathcal Q} |^2 e^{-\omega_{\textbf P} (\tau'-\tau )}  +\cdots \,,
\end{equation}
where, again, the ellipsis denotes the contribution from the multiparticle continuum. Here we have also introduced
\begin{equation}
Z_{\mathcal Q} \equiv \langle 0|  N_{\mathcal Q}(0)|\textbf P, \mathcal Q \rangle \,.
\end{equation}
From the Euclidean time dependence of this correlation function, one can isolate the energy of the particle as well as the matrix element of the interpolator. 
This has allowed the lattice QCD community to evaluate the QCD spectrum as well as state-to-vacuum matrix elements, which can be related, for example, to decay constants.

Similarly, matrix elements of operators between an incoming and outgoing single-particle state can be accessed using three-point functions of the form
\begin{equation}
C^{\mathcal O}_{\mathcal Q' \mathcal Q}(\tau', \tau, \{ \pmb \xi\} ,\textbf P', \textbf P) \equiv \langle  \widetilde N_{\mathcal Q'}(\tau', \textbf P')  
\mathcal O(0, \{\pmb \xi \})
   \widetilde N^\dagger_{\mathcal Q}(\tau, \textbf P) \rangle \,.
\label{eq:three_point}
\end{equation}
Inserting a complete set of states adjacent to both of the interpolating operators, one finds, 
\begin{align}
C^{\mathcal O}_{\mathcal Q' \mathcal Q}(\tau', \tau, \{ \pmb \xi \},  \textbf P', \textbf P)
& = 
\langle 0| N_{\mathcal Q'}(0)  |\textbf P', \mathcal Q '\rangle
\langle \textbf P', \mathcal Q' | \mathcal O(0, \{\pmb \xi \}) |\textbf P, \mathcal Q \rangle
\langle\textbf P, \mathcal Q | N^\dagger_{\mathcal Q}(0) |0 \rangle 
e^{-\omega_{\textbf P'} \tau'}
e^{\,\omega_{\textbf P} \tau}
\,+\cdots, \\
& = Z_{\mathcal Q'} Z_{\mathcal Q}^*
 \ \mathcal{M}_{\mathcal Q' \mathcal Q}^{\mathcal O} ( \{ \pmb \xi \}, \textbf P', \textbf P) \
e^{-\omega_{\textbf P'} \tau'}
e^{\,\omega_{\textbf P} \tau}
\,+\cdots \,,
\label{eq:three_point_v2}
\end{align}
with the ellipsis indicating the contribution of multiparticle states.  Extracting the leading-time behavior of this correlation function, and dividing out the interpolator matrix elements and time dependence, as determined from the two-point function, one can isolate the desired matrix element, Eq.~\eqref{eq:mat_elem}, directly from lattice QCD.
 
We conclude this section by reviewing how one extracts $\mathcal M_{\mathcal Q' \mathcal Q}^{\mathcal O}$ from a Minkowski correlator. We define
 \begin{equation}
C^{\mathcal O}_{M,\mathcal Q' \mathcal Q}( \{ \pmb \xi \},  P',  P) \equiv \int d^4 y' d^4 y \ e^{i P' y' - i P y} \langle  T  N_{\mathcal Q'}(y')  \mathcal O(0,\{\pmb \xi\})    N^\dagger_{\mathcal Q}(y) \rangle \,,
\end{equation}
where the subscript $M$ indicates that all four-momenta are defined with Minkowski signature and where $T$ indicates standard time ordering. In the on-shell limit, this correlator develops a pole for both the incoming and outgoing single particle states
\begin{equation}
C^{\mathcal O}_{M,\mathcal Q' \mathcal Q}( \{ \pmb \xi \},  P',  P) \sim \frac{i Z_{\mathcal Q'}}{P'^2 - \M_{\mathcal Q'}^2}         \mathcal{M}_{\mathcal Q' \mathcal Q}^{\mathcal O} ( \{ \pmb \xi \}, \textbf P', \textbf P)          \frac{i Z_{\mathcal Q}}{P^2 - \M_{\mathcal Q}^2} \,,
\end{equation}
where the $\sim$ indicates that the two sides differ by terms that are finite at the location of the combined poles. In a second step one must amputate the single particle propagators to access the desired matrix element. We do not describe this in any detail, because we only wish to point out that the matrix element accessed in this correlator is identically equal to that appearing in Eq.~\eqref{eq:three_point_v2}. The equality is simply definitional, since the same states and operators appear in the two cases.

\subsection{Operators displaced in time \label{sec:displaced}}

The analysis of the previous subsection hints as to why it is difficult to interpret matrix elements of operators that are evaluated at different Euclidean times. For example, for two currents displaced only in time, the corresponding matrix elements have a nontrivial time dependence given by
\begin{align}
\langle\textbf P, \mathcal Q|
  {\mathcal{J}}_E(\tau,0)
    {\mathcal{J}}(0)
    |\textbf P, \mathcal Q\rangle
=
\sum_{\mathcal Q'} \int \frac{d^3 \textbf k}{  (2\pi)^3  {2 \omega_{\textbf k}} } 
e^{-\tau(\omega_{\textbf k}-\omega_{\textbf P})}
\langle\textbf P, \mathcal Q|
  {\mathcal{J}}(0)
  |\textbf k,\mathcal Q'\rangle \langle \textbf k,\mathcal Q'|
    {\mathcal{J}}(0)
    |\textbf P, \mathcal Q\rangle+\cdots \,.
\end{align}
The Minkowski analogue, by contrast, takes the form
\begin{align}
\langle\textbf P, \mathcal Q|
  {\mathcal{J}}_M(t,0)
    {\mathcal{J}}(0)
    |\textbf P, \mathcal Q\rangle
=
\sum_{\mathcal Q'} \int \frac{d^3 \textbf k}{  (2\pi)^3 { 2 \omega_{\textbf k}}  } 
e^{-it(\omega_{\textbf k}-\omega_{\textbf P})}
\langle\textbf P, \mathcal Q|
  {\mathcal{J}}(0)
  |\textbf k,\mathcal Q'\rangle \langle \textbf k,\mathcal Q'|
    {\mathcal{J}}(0)
    |\textbf P, \mathcal Q\rangle+\cdots \,,
\end{align}
where in both cases the ellipsis indicates the contribution from multiparticle states. 

The relation between the two expressions is very complicated due to the integral over momentum. In a finite-volume this is replaced by a sum, but still the dependence on an infinite-tower of intermediate states obscures the relation between the matrix elements. Previous attempts to understand correlation functions of temporally displaced operators in systems with finite-volume Hamiltonians have been in the context of long-range effects in the calculations of $K_L-K_S$ mass difference~\cite{Christ:2015pwa}, and more work is needed to derive usable relations between Minkowski and Euclidean correlation functions for lattice applications.

\section{Quasi-distributions in perturbation theory\label{ssec:pertth}}

Reference~\cite{Carlson:2017gpk} highlights an apparent tension between the Minkowski and Euclidean evaluations of the quasi-PDF at one loop in perturbative QCD. The discrepancy arises because, for certain diagrams, the integration contours along the real and imaginary axes of the complex $k^0$ plane---$k$ denoting the integrated loop momentum---cannot be related by smooth deformation. The authors identify the result of integration along real $k^0$ as the physical observable and the quantity given by integrating along imaginary $k^0$ as the Euclidean object extracted in a lattice QCD calculation.

Although one cannot expect perturbation theory to capture the low-energy physics of PDFs, the discrepancy in the infrared behavior of these integrals presents a challenge to any calculation that relies on perturbation theory to relate quasidistributions directly to PDFs. More importantly, as we have argued in the previous section, there should be no distinction between the quasi-PDF extracted from Minkowksi and Euclidean correlators. Given this observation, it should be possible to implement any calculational scheme, regardless of its expected accuracy, in a way that gives the same prediction for the quasi-PDF, independent of the signature in the correlators.

In this section we examine this issue in the context of a toy model, focusing on a Feynman diagram that is directly analogous to that studied in Ref.~\cite{Carlson:2017gpk}. We first calculate the contribution from this diagram to the momentum-space Minkowski correlator and then use LSZ reduction to identify a contribution to our toy quasi-PDF that is directly analogous to that of Ref.~\cite{Carlson:2017gpk}. We find that the behavior of the corresponding momentum integral depends on whether one chooses the contour to lie on the real or the imaginary $k^0$ axis. However, we do not see any clear interpretation for the quantity defined with the second contour.

\begin{figure}
\begin{center}
{\includegraphics[width =.8\textwidth]{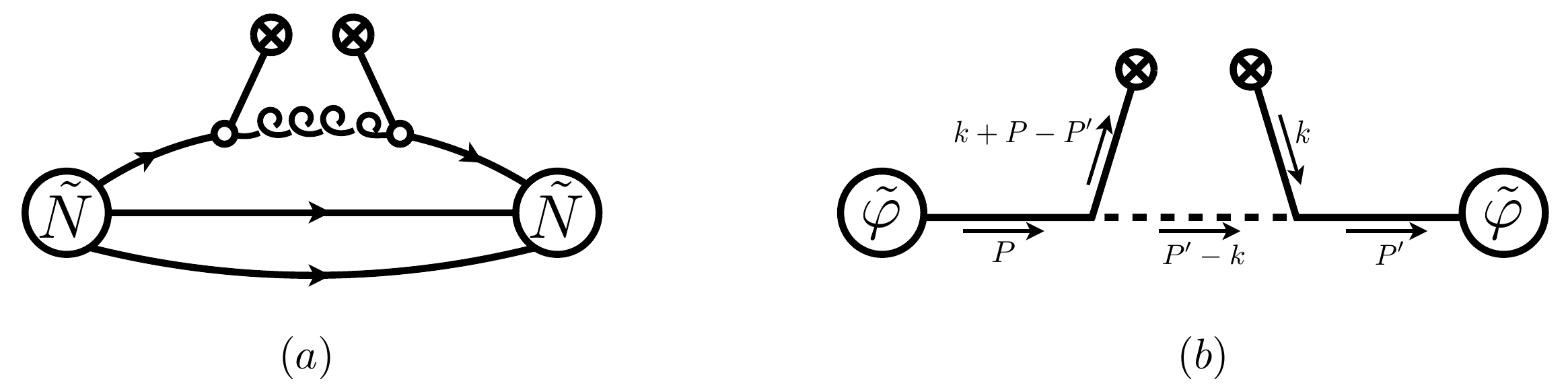} }
\caption{Perturbative contributions to the momentum-space correlation function involving a single insertion of a space-dislocated operator, depicted by the crossed circles, in (a) perturbative QCD and (b) the scalar toy model considered in the text. \label{fig:QCD_pert_PDF}}
\end{center}
\end{figure}

To make a clean connection to calculations performed in lattice QCD, we next compute the contribution of the same diagram to a Euclidean correlator in a mixed time-momentum representation. We then show that the large time limit is dominated by a term with the characteristic dependence of an initial and final single-particle state, and that the matrix element accompanying this time dependence is exactly that predicted by the Minkowski correlator. We consider a simplified model, because, although one can carry out this calculation in QCD, see Fig.~\ref{fig:QCD_pert_PDF}(a), the added complications associated with the full theory (the spin of the quarks and gluons, gauge invariance, and confinement) do not play any role in resolving the question at hand.

%%%%%%%%%%%%%%%%%%%% 
 %%%%%%%%%%%%%%%%%%%

We begin by introducing a pair of scalar fields:~$\varphi$, corresponding to a particle with physical mass $m$, and $\chi$, corresponding to a light degree of freedom with mass $m_\chi$. The limit in which $\chi$ becomes massless, $m_\chi \to 0$, most closely resembles the scenario considered in Ref.~\cite{Carlson:2017gpk}. We introduce a nonzero mass to avoid complications from the fact that the single particle pole will coincide with the $\varphi+\chi$ production threshold. As will become clear shortly, the observations made in Ref.~\cite{Carlson:2017gpk} hold even with a nonvanishing value for $m_\chi$.  

We define the action of this theory to include the interaction term
\begin{equation}
S_{\varphi^2 \chi} \equiv  \frac{g}{2} \int d^4 x \varphi(x)^2 \chi(x) \,,
\end{equation}
where $g$ is a coupling constant. We then define a toy quasi-PDF
\begin{equation}
\label{eq:qPDF}
\widetilde q(x,P_z) \equiv \int d \xi_z e^{i \xi_z x P_z}   \langle \textbf P \vert  \varphi(\xi) \varphi(0)   \vert \textbf P \rangle \,,
\end{equation}
where $\vert \textbf P \rangle$ is the single-particle state interpolated by $\varphi$ with three-momentum $\textbf P = (0,0,P_z)$ and $\xi$ is a four-vector in which all components vanish aside from the third spatial component, $\xi_z$. Here we use the mostly-minus metric with $\xi^3 = - \xi_3$ but also follow the convention that a variable with a $z$ subscript corresponds to an upper-index coordinate, e.g.~$\xi^\mu = (0,0,0,\xi_z)$.  

Since the matrix element in Eq.~\eqref{eq:qPDF} depends only on Hamiltonian eigenstates and operators with vanishing time components, it cannot be assigned the label of Minkowski or Euclidean. The matrix element is, however, very simply related to a Minkowski correlator defined by
\begin{equation}
C_M(x, P', P) \equiv  \int d \xi_z e^{i \xi_z x P_z}   \int d^4  y'  \int d^4 y  e^{- i  P'  y'}  e^{+ i  P  y}  \langle 0 \vert  \varphi(y')      \left [    \varphi(\xi) \varphi(0)       \right ]      \varphi(y)  \vert 0 \rangle \,.
\end{equation}
From the LSZ reduction formula follows
\begin{equation}
\label{eq:LSZ}
\lim_{P^0 \to \omega_{\textbf P}, P'^0 \to \omega_{\textbf P}} i^2 (P^2 - m^2) (P'^2 - m^2) C_M(x, P', P) = \widetilde q(x, P_z) \,,
\end{equation}
where $\omega_{\textbf P}\equiv \sqrt{P_z^2 + m^2}$. Note also that we have normalized the fields so that the $\varphi$ propagator has residue 1 at the pole. Equation \eqref{eq:LSZ} is simply the statement that the quasi-PDF is the residue of the Minkowski correlator at the combined incoming and outgoing single-particle poles:
\begin{equation}
C_M(x, P', P) \sim \frac{i}{P'^2 - m^2}    \widetilde q(x,P_z)     \frac{i}{P^2 - m^2} \,.
\end{equation}

Our aim is consider a particular perturbative contribution to $C_M$ and the corresponding contribution to $\widetilde q(x,P_z)$. It is convenient to first rewrite the correlator as 
\begin{equation}
C_M(x,P', P) =   \int \frac{d k^1 d k^2 d k^0}{(2 \pi)^3}       \langle 0 \vert \widetilde \varphi(P')       \widetilde   \varphi(- k) \varphi(0)     \widetilde        \varphi(-P)  \vert 0 \rangle \,,
\end{equation}
where $k^\mu =(k^0, k^1,k^2, x P_z)$, $P^\mu = (P^0,0,0,P_z)$ and $P'^\mu = (P'^0,0,0,P_z)$. The momentum-dependent four-point function appearing here is precisely the quantity used to calculate two-to-two scattering of incoming particles with momenta $k$ and $P$ to outgoing particles with $P'$ and $k+P-P'$. Since one of the $\varphi$ fields is not projected to definite momentum, this correlator is not proportional to a momentum-conserving delta function. Of course, one must still enforce momentum conservation in Feynman diagrams to correctly calculate this quantity.

We are now ready to evaluate the contribution to this correlator from the diagram shown in Fig.~\ref{fig:QCD_pert_PDF}(b). The calculation is straightforward for this tree-level diagram and we find
\begin{align}
C^{(1)}_M(x,P', P) = {} & (ig)^2 \frac{i}{P'^2 - m^2 + i \epsilon} \nonumber \\
{} & \times \left[  \int \frac{d k^1 d k^2 d k^0}{(2 \pi)^3}     \frac{i}{k^2 - m^2 + i \epsilon} \frac{i}{(P'-k)^2-m_\chi^2 + i \epsilon} \frac{i}{(k+P-P')^2 - m^2 + i \epsilon}  \right ]  \frac{i}{P^2 - m^2 + i \epsilon} \,,
\end{align}
where the superscript $(1)$ indicates we are considering the contribution only from the diagram shown in the figure. From Eq.~\eqref{eq:LSZ} then follows
\begin{equation}
\widetilde q^{(1)}(x,P_z) =   g^2 I_M(x,P_z) \,,
\end{equation}
where
\begin{equation}
I_M(x,P_z) \equiv    i       \int \frac{d k ^0 d k^1 d k^2}{(2 \pi)^3}   \left ( \frac{1}{k^2 - m^2 + i \epsilon}  \right )^2 \frac{1}{(P-k)^2 -m_\chi^2+ i \epsilon} \ \bigg \vert_{P^\mu = (\omega_{\textbf P},0,0,P_z) , \ k^\mu = (k^0,k^1,k^2,xP_z)} \,.
\end{equation}
In the context of this toy model, this represents the desired, physical matrix element.

\bigskip

We now aim to understand what one would extract in a numerical lattice QCD calculation in this toy theory, if it were dominated by this diagram. Given that such a calculation can only access Euclidean correlators, it is natural to expect the calculated quantity might be
\begin{equation}
\widetilde q^{(1)}_E(x,P_z) =  g^2 I_E(x,P_z) \,,
\end{equation}
where
\begin{equation}
I_E(x,P_z) \equiv     \int \frac{d k_4 d k_1 d k_2}{(2 \pi)^3}   \left ( \frac{1}{k^2 + m^2 }  \right )^2 \frac{1}{(P-k)^2 +m_\chi^2} \ \bigg \vert_{P_\mu = (0,0,P_z,i\omega_{\textbf P}) , \ k_\mu = (k_1,k_2,xP_z, k_4)} \,.
\end{equation}
We stress that all four-vectors in this expression are defined with Euclidean convention, i.e.~$k^2 = \sum_{\mu=1}^4 k_\mu k_\mu$. Note also that we have set $P_4 = i \omega_{\textbf P}$ in order to preserve the on-shell condition defining the quasi-PDF.

%%%%%%%%%%%%%%%%%%%%
\begin{figure}
\begin{center}
\includegraphics[width =.8\textwidth]{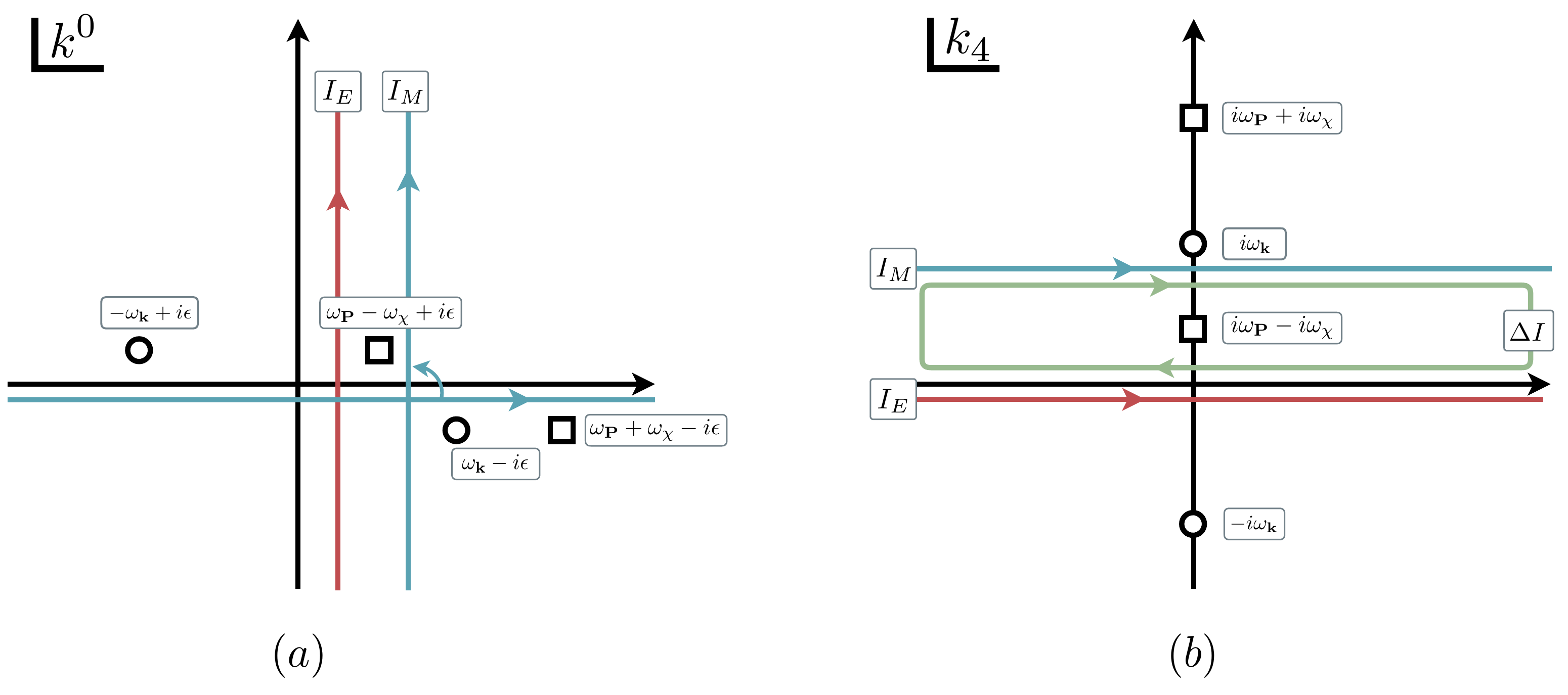}
\caption{Analytic structure of the integrand in the (a) Minkowski and (b) Euclidean coordinate space, together with contours for the $I_M$, $I_E$, and $\Delta I$ integrals discussed in the text.\label{fig:contour}}
\end{center}
\end{figure}

%%%%%%%%%%%%%%%%%%%%

As was pointed out in Ref.~\cite{Carlson:2017gpk}---in that work using a slightly different integrand---$I_M \neq I_E$. This can be seen by recognizing that $I_M$ and $I_E$ can be expressed as integrals over the same integrand, but along different contours in the complex $k_0$, or $k_4$, plane. This difference was explained in Ref.~\cite{Carlson:2017gpk}, and we summarize their discussion for convenience. As we illustrate in Fig.~\ref{fig:contour}(a), the integrals differ because, for $k_1$ and $k_2$ satisfying
\begin{equation}
k_1^2 + k_2^2 < \mu^2 \equiv P_z^2 - (1-x)^2 P_z^2 + m^2 -m_\chi^2 \,,
\end{equation}
the contours defining $I_E$ and $I_M$ are separated by a pole at
\begin{equation}
\label{eq:kapdef}
k_4 = i \kappa \equiv  i  \omega_{\textbf P}- i \omega_\chi  \,,
\end{equation}
where
\begin{align}
\omega_\chi&\equiv\sqrt{m_\chi^2+\vert \textbf P - \textbf k \vert^2} \,, \\
\vert \textbf P - \textbf k \vert &\equiv \sqrt{k_1^2 + k_2^2 + (1 - x)^2 P_z^2} \,.
\end{align}
Thus, for $k_1^2 + k_2^2<\mu^2$, an attempt to deform the $k_4$ contour defining $I_E$ into the contour defining $I_M$ leads to an additional piece, encircling the pole at $k_4 = i \kappa$. Integrating the residue over  $k_1^2 + k_2^2<\mu^2$ then gives the nonzero difference between the two integrals.

This series of observations nicely illustrates potential subtleties in relating Minkowski and Euclidean quantities. Given that $I_E \neq I_M$, the claim that $I_E$ would be extracted on the lattice contradicts the statement that such a calculation must yield the physical quantity. The resolution is that the prescription used to define $I_E$ does not correctly represent how the quasi-PDF is extracted from a lattice calculation. The correct approach is to study a Euclidean correlator in a mixed time-momentum representation and to identify the leading large time-dependence, associated with the single-particle matrix element. As we now demonstrate, the matrix element extracted in this way, from a strictly Euclidean correlator, is the physically-relevant and expected result, $I_M$.

\bigskip

Before focusing on the diagram of Fig.~\ref{fig:QCD_pert_PDF}(b), we recall that in a lattice calculation of this toy theory, one directly accesses a Euclidean correlator of the form  \begin{equation}
C_E(\tau', \tau, x, P_z) \equiv \int d \xi_z e^{i \xi_z x P_z}   \int d^3 \textbf y'  \int d^3 \textbf y  e^{- i  P_z  y'_3}  e^{+ i  P_z  y_3}  \langle 0 \vert  \varphi(y')      \left [    \varphi(\xi) \varphi(0)       \right ]      \varphi(y)  \vert 0 \rangle \,,
\end{equation}
where $\varphi(y)$ creates a particle at early times, $\varphi(y')$ annihilates a particle at late times and the product in square brackets is our toy quasi-PDF operator. The relevant four-vectors are given by $\xi_\mu = (0, 0, \xi_z, 0)$, $y_\mu=(\textbf y, \tau)$ and $y_\mu'=(\textbf y', \tau')$, and we assume $\tau < 0 < \tau' $ throughout. We stress that all four-vectors used in the remainder of this section have Euclidean signature. As above, it is convenient to express $C_E$ in terms of a purely momentum-space four-point function
\begin{equation}
C_E(\tau', \tau, x, P_z) =   \int \frac{d P_4'}{2 \pi}       e^{  i P_4' \tau'  }    \int \frac{d P_4}{2 \pi}   e^{ - i P_4 \tau  }     \int \frac{d k_1 d k_2 d k_4}{(2 \pi)^3}           \langle 0 \vert \widetilde \varphi(P')       \widetilde   \varphi(- k) \varphi(0)     \widetilde        \varphi(-P)  \vert 0 \rangle \,,
\end{equation}
where $k_\mu =(k_1,k_2, x P_z, k_4)$, $P_\mu = (0,0,P_z,P_4)$ and $P_\mu' = (0,0,P_z,P_4')$.

We are ready to return to the contribution from the diagram in Fig.~\ref{fig:QCD_pert_PDF}(b)
\begin{align}
C^{(1)}_E(\tau', \tau, x, P_z) \equiv  {} & g^2   \int \frac{d k_1 d k_2 d k_4}{(2 \pi)^3}         \int \frac{d P_4'}{2 \pi}       e^{  i P_4' \tau'  }  \int \frac{d P_4}{2 \pi}   e^{ - i P_4 \tau  } \nonumber \\ 
{} & \qquad \times  \frac{1}{P_4'^2 + P_z^2 + m^2}  \frac{1}{k^2 + m^2} \frac{1}{(P'-k)^2+m_\chi^2} \frac{1}{(k+P-P')^2 + m^2} \frac{1}{P_4^2 + P_z^2 + m^2} \,.
\end{align}
We stress that the $k_4$ integral is evaluated along the real axis, and only deformations of the contour that do not cross poles may be performed. Thus, at this stage the correlator appears more closely related to $I_E$ than to $I_M$. To make a clean comparison, we must now evaluate the $P_4$ and $P_4'$ integrals and then pick off the leading time dependence, associated with the single-particle initial and final states. In particular we aim to identify the term that scales as $e^{- \omega_{\textbf P}(\tau' - \tau)}$. Since we have a mass gap separating the single-particle pole from the $\varphi+\chi$ threshold, this is the dominant term in the large $ (\tau' - \tau  )$ limit.

Given that $\tau < 0$ and $\tau' >0$, together with the form of the exponentials, we see that both the $P_4$ and $P_4'$ integrals should be evaluated by closing the contours in the upper half of their respective complex planes. Beginning with the $P_4$ integral, note that this can encircle either the pole at $P_4 = i \omega_{\textbf P}$ or else at $P_4 = i \omega_{\textbf k} +P_4' - k_4$, where we have introduced
\begin{equation}
\omega_{\textbf k} \equiv \sqrt{k_1^2 + k_2^2 + x^2 P_z^2 + m^2 } \,.
\end{equation}
The sum of these two contributions gives
\begin{align}
C^{(1)}_E{} & (\tau', \tau, x, P_z) \equiv  g^2 \frac{e^{\omega_{\textbf P} \tau}}{2 \omega_{\textbf P}}   \int \frac{d k_1 d k_2 d k_4}{(2 \pi)^3}     \int \frac{d P_4'}{2 \pi}       e^{  i P_4' \tau'  }    \frac{1}{P_4'^2 + \omega_{\textbf P}^2}  \frac{1}{k^2 + m^2} \frac{1}{(P'-k)^2+m_\chi^2} 
\frac{1}{(k_4+i\omega_{\textbf P}-P'_4)^2 + \omega_{\textbf k}^2}   \nonumber \\
{} & + g^2  \int \frac{d k_1 d k_2 d k_4}{(2 \pi)^3}   \frac{1}{2 \omega_{\textbf k}}       \int \frac{d P_4'}{2 \pi}      e^{ - i (i \omega_{\textbf k}  - k_4)  \tau  }     e^{  i P_4' ( \tau'  - \tau)}    \frac{1}{P_4'^2 + \omega_{\textbf P}^2}  \frac{1}{k^2 + m^2} \frac{1}{(P'-k)^2+m_\chi^2}   \frac{1}{( i \omega_{\textbf k} +P_4' - k_4)^2 + \omega_{\textbf P}^2} \,.
\end{align}

 At this stage there are a total of seven poles that can potentially appear in the upper-half of the complex $P_4'$ plane. In the first term these are
\begin{equation}
P'^A_4 = i \omega_{\textbf P} \,, \ \ \  P'^B_4 = i \omega_\chi + k_4 \,, \ \ \  P'^C_4 = k_4 +  i  \omega_{\textbf P} + i \omega_{\textbf k} \,,  \ \ \  P'^D_4 = k_4  - i \omega_{\textbf k} +  i  \omega_{\textbf P} \,,
\end{equation}
and in the second term
\begin{equation}
P'^E_4 = i \omega_{\textbf P} \,, \ \ \  P'^F_4 = i \omega_\chi + k_4 \,, \ \ \  P'^G_4 = k_4 - i \omega_{\textbf k} + i \omega_{\textbf P} \,.
\end{equation}
A careful analysis shows that the poles labeled $C$, $E$ and $F$ give no contribution to the time-dependence we are after. In addition, although the poles labeled $D$ and $G$ individually give relevant time dependence, the contributions cancel identically between the two terms.

Thus the only relevant terms come from the poles labeled $A$ and $B$. Evaluating the $P_4'$ integral by encircling these two gives
\begin{align}
C^{(1)}_E(\tau', \tau, x, P_z) \equiv {} &  g^2 \frac{e^{- \omega_{\textbf P} ( \tau' - \tau)}}{4 \omega_{\textbf P}^2}   \int \frac{d k_1 d k_2 d k_4}{(2 \pi)^3}    \left( \frac{1}{k^2 + m^2} \right )^2 \frac{1}{(P-k)^2+m_\chi^2}\  \bigg \vert_{P_\mu = (0,0,P_z,i\omega_{\textbf P}) , \ k_\mu = (k_1,k_2,xP_z, k_4)} \nonumber \\
{} & +g^2 \frac{e^{\omega_{\textbf P} \tau}}{2 \omega_{\textbf P}}   \int \frac{d k_1 d k_2 d k_4}{(2 \pi)^3}      \frac{    e^{  i (i \omega_\chi + k_4) \tau'  } }{2 \omega_\chi}   \frac{1}{(i \omega_\chi + k_4)^2 + \omega_{\textbf P}^2}  \frac{1}{k^2 + m^2}  \frac{1}{( i \omega_{\textbf P}   -   i \omega_\chi )^2 + \textbf k^2 + m^2}   + \cdots  \,,
\end{align}
where the ellipsis stands for terms that are suppressed for large source-sink separation.

To complete our discussion we substitute the definition for $I_E$ into the first term. In addition, we rearrange the second term using $\kappa$, defined in Eq.~\eqref{eq:kapdef} above
\begin{align}
C^{(1)}_E(\tau', \tau, x, P_z) \equiv  {} & g^2 \frac{e^{- \omega_{\textbf P} ( \tau' - \tau)}}{4 \omega_{\textbf P}^2}  \nonumber \\ 
{} & \times \bigg (  I_E(x, P_z) 
+   \int \frac{d k_1 d k_2 d k_4}{(2 \pi)^3}       \frac{    e^{  i (  k_4 - i \kappa) \tau'  } }{k_4 - i \kappa}   \frac{2 \omega_{\textbf P}}{2 (   k_4 + i \omega_\chi)    }  \frac{1}{2 \omega_\chi} \frac{1}{k^2 + m^2} \left [ \frac{1}{k^2 + m^2}\bigg \vert_{k = i \kappa} \right ]  \bigg ) + \cdots  \,.
\end{align}
From this result one can clearly see that the integral around $k_4 = i \kappa$ generates a time-independent contribution to the second term in parenthesis. Keeping only this contribution, we reach our final result
\begin{equation}
C^{(1)}_E(\tau', \tau, x, P_z) \equiv  g^2 \frac{e^{- \omega_{\textbf P} ( \tau' - \tau)}}{4 \omega_{\textbf P}^2} [ I_E(x, P_z) + \Delta I (x,P_z) ] + \cdots  \,,
\end{equation}
where
\begin{align}
\Delta I(x, P_4) \equiv & {}   \int \frac{d k_1 d k_2  }{(2 \pi)^2}  \Theta(\mu^2 - k_1^2 - k_2^2)       \frac{1}{2 \omega_\chi}    \left ( \frac{1}{k^2 + m^2} \right )^2_{k_4 = i \kappa} \nonumber \,, \\  = & {}  \int \frac{d k_1 d k_2  }{(2 \pi)^2} \Theta(\mu^2 - k_1^2 - k_2^2)  \ointclockwise_{i \kappa}  \frac{d k_4}{2 \pi}    \left (  \frac{1}{k^2 + m^2}  \right )^2 \frac{1}{(P-k)^2+m_\chi^2} \ \bigg \vert_{P_\mu = (0,0,P_z,i\omega_{\textbf P}) , \ k_\mu = (k_1,k_2,xP_z, k_4)}   \,.
\end{align}
Here the theta function is required because this contribution only arises when $\kappa>0$.
In the second step we have noted that $\Delta I$ can be written as the clockwise contour integral, around the $i \kappa$ pole, of the same Euclidean integrand defining $I_E$.

As is apparent from Fig.~\ref{fig:contour}(b), $\Delta I $ is precisely the term needed to convert $I_E$ to $I_M$. Thus we find that, although the contour integral around the external poles leads to a term that depends only on $I_E$, a careful evaluation of the full diagram ensures that the lattice calculation will indeed extract $I_M$, as it must. 

\begin{figure}
 \begin{center}
 \includegraphics[width =.8\textwidth]{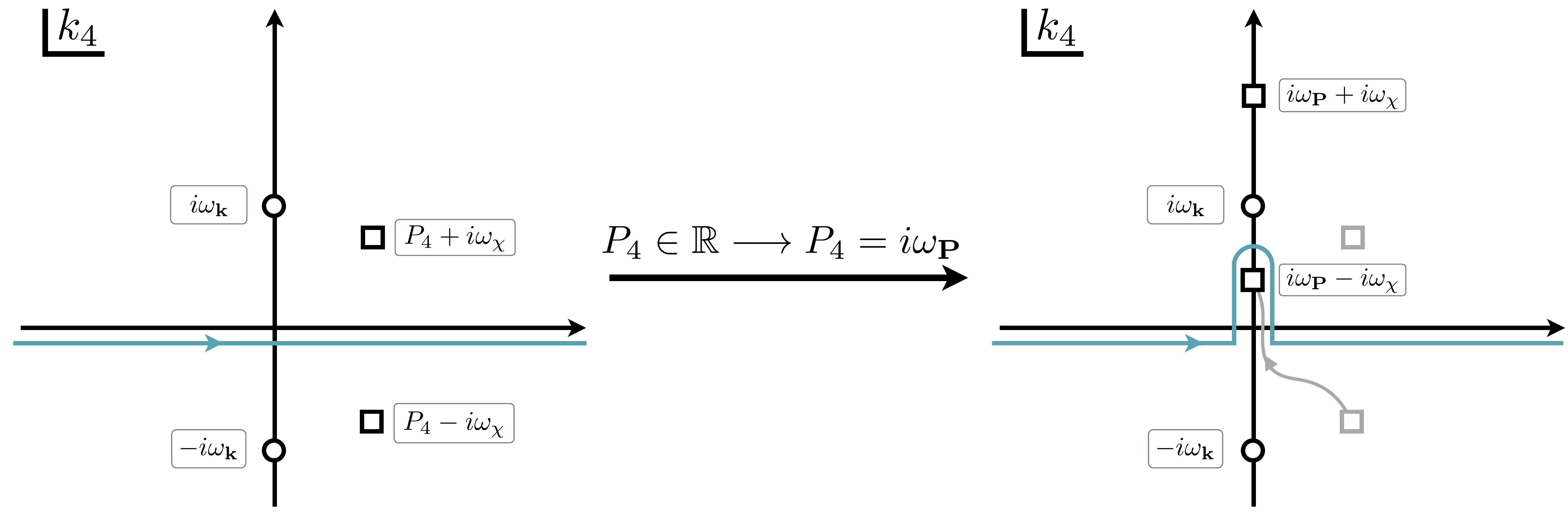}
 \vspace{-10pt}
 \caption{Analytic continuation of the momentum-space Euclidean correlator from real $P_4$ (left) to the on-shell point (right).  }\label{fig:fromRealP} 
 \end{center}
 \end{figure}

 To gain further insight into the recovery of $I_M$, it is instructive to consider the complex $k_4$ plane of the integrand for real values of $P_4$, shown in Fig.~\ref{fig:fromRealP}. When $P_4$ is real the on-shell condition cannot be satisfied, but the off-shell Euclidean correlator is unambiguously defined by integration along the real $k_4$ axis. The key point is that, rather than setting $P_4 = i \omega_{\textbf P}$ in the integrand, one must analytically continue the integral from real to on-shell $P_4$. This analytic continuation is demanded by the calculation of the Euclidean-time-dependent correlator. The Fourier transform is initially defined only on the real $P_4$ axis and then one continues into the complex plane and encircles poles as a tool for evaluating the integral. In this sense the recipe of continuing from real $P_4$ is built into a lattice calculation.
 
As we show in Fig.~\ref{fig:fromRealP}, analytic continuation of the integral is effected by smoothly deforming the integrand while insisting that poles can never cross contours. In the present case the pole crosses the real axis as $P_4$ approaches the on-shell point, and thus the contour must be deformed into the complex plane. This leads to a definition for the correctly continued correlator that corresponds to $I_M$ rather than $I_E$, as expected. This perspective also gives us a general recipe for identifying the problematic diagrams: If, for a given integrand, the pole crosses the real $k_4$ axis as $P_4$ is varied from real values to the on-shell point, then the contour must be deformed from the real $k_4$ axis to extract the lattice prediction of the diagram.\footnote{ We note that this prescription has previously been applied to perturbative calculations of local heavy-light currents \cite{Hart:2006ij}.}

In the following section we prove, to all orders in perturbation theory, that the result of integrating along this properly deformed contour is identically equal to the Minkowski definition of the same diagram.

\section{ All-orders uniqueness proof \label{ssec:all_orders}}
In the previous section we investigated a particular perturbative contribution to a Euclidean three-point function and demonstrated that the contribution to the matrix element is equal to the result obtained via a Minkowski correlator. Here we give an all-orders perturbative proof of this equality, that holds for any field theory with a mass gap between the single-particle pole and the first threshold. This derivation also requires that the zero-time current insertion can be expressed as an infinite series of local operators built from the low-energy degrees of freedom of the theory. We comment that the correspondence shown here is implicit in various all-orders studies of finite-volume Euclidean correlation functions involving electroweak currents (see, for example, Refs.~\cite{Bernard:2012bi, Briceno:2014uqa, Agadjanov:2014kha, Briceno:2015tza, Briceno:2015csa}).

We first recall the correlator definitions of Sec.~\ref{sec:euclidean}. Here, to avoid clutter of notation, we drop the $\mathcal Q'$ and $\mathcal Q$ labels, indicating general quantum numbers. In Eqs.~\eqref{eq:three_point}-\eqref{eq:three_point_v2} we introduced a class of Euclidean three-point functions and picked off the leading time dependence
\begin{equation}
\label{eq:Etime}
C_E(\tau', \tau, \{\textbf Q \} , \textbf P', \textbf P) \equiv  \langle    N(\tau',\textbf{P}')  \widetilde {\mathcal O}(\{\textbf Q\})    N^\dagger(\tau,\textbf{P}) \rangle = 
 \mathcal{M}_E^{\mathcal O} (\{\textbf Q \}, \textbf P', \textbf P)
\frac{e^{-\omega_{\textbf P'}  \tau' + \omega_{\textbf P} \tau }}{4 \omega_{\textbf P} \omega_{\textbf P'}}
\,+\cdots \,.
\end{equation}
In contrast to Sec.~\ref{sec:euclidean}, here we have simplified the expressions by assuming the interpolator $N$ satisfies $ \langle 0| N(0)  |\textbf P \rangle = 1$. We have also given an $E$ subscript to the coefficient of the single-particle exponential, to indicate that it is extracted from a Euclidean correlation function. 

We allow the current insertion, $\widetilde {\mathcal O}(\{\textbf Q\})$, to depend on any number of spatial momenta $\{\textbf Q\} = \{\textbf Q_1, \cdots , \textbf Q_m\}$. In the low-energy effective theory, the current is generally represented as a sum of terms each built from products of the single-particle interpolators corresponding to low-energy degrees of freedom. We allow the spatial structure of these interpolators to be arbitrary, and note that a general term can be expressed as a product of fields at definite spatial-momenta, possibly with momentum integrals and momentum dependent weight functions. We do, however, insist that the time components of all fields are evaluated at $\tau=0$, i.e.~that the operator is not displaced in time.

 In Sec.~\ref{sec:euclidean} we argued on general grounds that this coefficient must equal the physical matrix element
\begin{equation}
{\mathcal M}^{\mathcal O}(\{\textbf Q \},\textbf P', \textbf P)  \equiv \langle \textbf P' | \widetilde{ \mathcal O}(\{\textbf Q\}) |\textbf P \rangle \,.
\end{equation}
In this section we prove that this claim holds to all orders in perturbation theory. More concretely, we define $ \mathcal{M}_E^{\mathcal O,(d)} (\{\textbf Q \}, \textbf P', \textbf P)$ as the contribution to the coefficient in Eq.~\eqref{eq:Etime}, from a generic diagram, labeled $d$. Similarly, we denote by $\mathcal M^{\mathcal O, (d)}(\{\textbf Q \}, \textbf P', \textbf P)$ the contribution from $d$ to the physical matrix element, as extracted from a Minkowski correlator. We then demonstrate 
\begin{equation}
\label{eq:proof}
\mathcal M_E^{\mathcal O, (d)}(\{\textbf Q \}, \textbf P', \textbf P) = \mathcal M^{\mathcal O, (d)}(\{\textbf Q \}, \textbf P', \textbf P)\,.
\end{equation}

Beginning with Eq.~\eqref{eq:Etime}, we first express the time-dependent correlation function as a Fourier transform of a pure momentum-space Euclidean correlator
\begin{align}
\label{eq:Etrans}
C^{(d)}_E(\tau', \tau, \{\textbf Q\}, \textbf P', \textbf P) 
=
\int \frac{d P_4'}{2 \pi}       e^{  i P_4' \tau'  }  \int \frac{d P_4}{2 \pi}   e^{ - i P_4 \tau  }
C^{(d)}_{E}(\{\textbf Q\},P',P) \,.
\end{align}
Combining Eqs.~\eqref{eq:Etime} and \eqref{eq:Etrans} then gives
\begin{align}
C^{(d)}_{E}(\{\textbf Q\},P',P)
\sim
\frac{1}{P'^2 + m^2}
{\mathcal M}_E^{\mathcal O, (d)}(\{\textbf Q\},\textbf P',\textbf{P})
 \frac{1}{P^2 + m^2},
 \label{eq:CE_at_poles2} \,
\end{align}
where $\sim$ indicates that the two sides differ by analytic terms near the two single-particle poles. 
Note that $\mathcal M_E^{\mathcal O}$ has no $P_4$ or $P_4'$ dependence, so the contour integral trivially gives this object as the coefficient of the single-particle exponential.   As mentioned above, here it is understood that the pair of single-particle poles is determined via analytic continuation from real $P_4$ and $P_4'$.

We next note that the contribution of diagram $d$ to $C_E(\{\textbf Q\},P',P)$ has the following general form
\begin{align}
\label{eq:Ediag}
C^{(d)}_E(\{\textbf Q\},P',P) = {} &  
\frac{1}{P'^2 + m^2} \int \frac{d^4k_{1}}{(2\pi)^4}
\frac{d^4k_{2}}{(2\pi)^4}\cdots \frac{d^4k_{n}}{(2\pi)^4}
\,
f^{(d)}_E(\{\textbf Q\},P', P, k_{1},k_{2},\ldots, k_{n}) \nonumber \\ 
{} & \qquad \times
\,
\left[\frac{1}{q^2_1+m_1^2}\frac{1}{q^2_2+m_2^2}\cdots\frac{1}{q^2_{n'}+m_{n'}^2}\right] \frac{1}{P^2 + m^2} \,,
\end{align}
 where $\{k_1,k_2,\ldots,k_n\}$ denote the integrated loop momenta and the $n'$ internal propagators depend on various linear combinations of the external and loop momenta, denoted $\{q_1,q_2,\ldots,q_{n'}\}$, as well as the various particle masses, $\{m_1,m_2,\ldots,m_{n'}\}$. The function $f_E$ contains symmetry factors, couplings, and weight-functions from the current insertion, leading to various polynomials in the indicated momentum coordinates. To reach this expression one must write the particular low-energy operator contributing to $\widetilde {\mathcal O}(\{\textbf Q\})$ with all single-particle fields at definite momenta. Then Eq.~\eqref{eq:Ediag} accommodates all possible operator structures, as long as we allow $f_E$ to also contain three dimensional delta functions to remove integrals over externally projected momenta. We stress that, in this expression, all integrated four-momenta as well as the external four-momenta, $P'$ and $P$, are defined as real Euclidean four-vectors. 
 
 Our aim at this stage is to analytically continue this expression from real $P_4$ and $P_4'$ to the on-shell values $P_4 = i \omega_{\textbf P}$ and $ P_4' = i \omega_{\textbf P'}$, and so determine the value of the residue factor, $\mathcal M_E^{\mathcal O, (d)}(\{\textbf Q\},\textbf P',\textbf P)$, associated with the diagram. In principle this continuation can be achieved by analytically varying $P_4$ and $P_4'$ away from the real axis, while maintaining the Euclidean conventions for all loop-momenta. However, as we have seen from the example of the previous section, a subtlety arises when the continuation of $P_4$ or $P_4'$ pushes a pole over the real axis of any of the $k_{4,n}$ integrals. Keeping the $k_{4,n}$ contour fixed and allowing the pole to cross that contour leads to a new function that differs from the analytic continuation that we are after. Instead one must deform the contour into the complex plane to prevent the crossing, as we described at the end of the previous section.
 
 For the general diagram considered here, it is very tedious to keep track of all possible crossings and the associated contour deformations. To avoid this, we instead pursue a global Wick rotation of all integrated coordinates as well as the external coordinates. In this approach one can generally argue that poles never cross contours and thus that the final result is the desired unique analytic continuation to the on-shell pole. We emphasize that this analytic continuation is automatically sampled in the Fourier transform leading to $C_E(\tau', \tau, \{\textbf Q\},\textbf P',\textbf P)$ and is therefore the correct prescription for accessing the quantity that is determined on the lattice.
 
\begin{figure}
 \begin{center}
 \includegraphics[width =.9\textwidth]{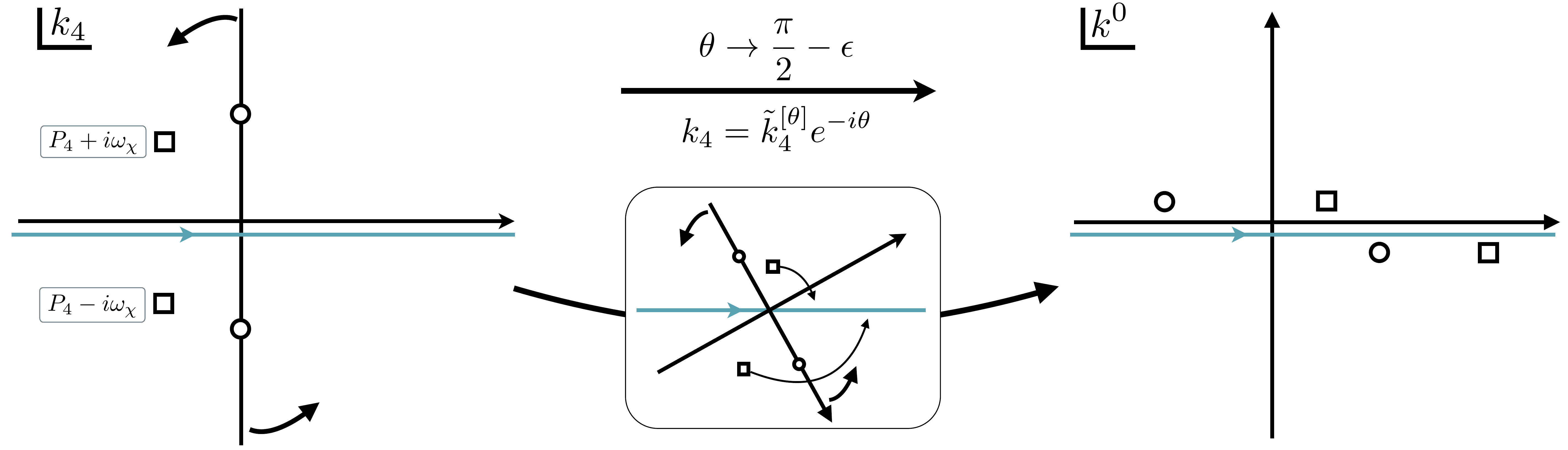}
 \caption{An example of the analytic continuation of the fourth component of momentum from Euclidean to Minkowski, discussed in the text. The left-hand panel shows the structure of the integrand for some negative, real value of $P_4$. The Wick rotation then is effected by changing the coordinate system while keeping the contour along the real $\widetilde k_4^{[\theta]}$ axis. The smaller middle panel shows the original $k_4$ axes midway through the rotation. As indicated by the arrows attached to the square poles in the middle panel, $P_4$ is continued to imaginary values simultaneously as the contour rotates in a way that ensures that poles never cross the contour. The final result, shown in the right-hand panel, is a Minkowski signature integral with pole locations satisfying the standard $i \epsilon$ prescription.} \label{fig:wick} 
 \end{center}
 \end{figure}

 To perform the rotation we define the coordinate transformations [see Fig.~\ref{fig:wick}]
 \begin{equation}
 k_{4,j} \equiv \widetilde k_{4,j}^{[\theta]} e^{-i \theta} \,, \ \ P_{4} \equiv \widetilde P_4^{[\theta]} e^{-i \theta} \,, \ \ P_4' \equiv \widetilde P_4'^{[\theta]} e^{-i \theta}  \,.
 \end{equation}
 We then note that the denominator of a generic propagator takes the form
 \begin{equation}
 q_{4,j}^2 + \omega_{\textbf q_j}^2 = ( \widetilde q_{4,j}^{[\theta]}  e^{-i \theta} - i \omega_{\textbf q_j} ) ( \widetilde q_{4,j}^{[\theta]} e^{-i \theta} + i \omega_{\textbf q_j} ) \,.
 \end{equation}
 From this expression it is manifest that, if one varies $\theta$ from $0$ to $\pi/2 - \epsilon$ while keeping $\widetilde k_j^{[\theta]}$, $\widetilde P_4^{[\theta]}$ and $\widetilde P_4'^{[\theta]}$ real, then all propagator denominators remain nonzero and so a pole never crosses a contour. Thus, this transformation, in which the external coordinates are varied as the internal coordinates are rotated, gives a general prescription for analytically continuing to the on-shell pole.
 
 The result of the rotation is then 
  \begin{align}
C^{(d)}_E(\{\textbf Q\},P',P) \sim {} & \frac{1}{P_4'^2 + \omega_{\textbf P'}^2} (-1)^{n'} (-i)^n \int \frac{d^4\widetilde k_{1}}{(2\pi)^4} \frac{d^4\widetilde k_{2}}{(2\pi)^4}\cdots \frac{d^4\widetilde k_{n}}{(2\pi)^4} 
\,
f_M(\{\textbf Q\},\widetilde P', \widetilde P,\widetilde k_{1},\widetilde k_{2},\ldots, \widetilde k_{n}) \nonumber \\ 
{} & \qquad \times
\,
\left[\frac{1}{{\widetilde q}^2_1-m^2_1+i\epsilon}\frac{1}{{\widetilde q}^2_2-m^2_2+i\epsilon}\cdots\frac{1}{{\widetilde q}^2_{n'}-m^2_{n'}+i\epsilon}\right]  \frac{1}{P_4^2 + \omega_{\textbf P}^2} \,,
\end{align}
where the $\sim$ indicates that the sides are equal at the pair of single-particle poles, and where $\widetilde P$, $\widetilde k_j$ and $\widetilde q_j$ are Minkowski signature four-vectors. To make contact with Eq.~\eqref{eq:CE_at_poles2} we have expressed the external propagators in terms of $P_4 = i \widetilde P^0$. Although we use the Euclidean convention, we stress that the expression has been analytically continued and the new result is no longer defined on the real axis, but instead defined near $P_4= i \omega_{\textbf P}$ and $P_4' = i \omega_{\textbf P'}$.\footnote{Since the rotation is given by $P_4 = \widetilde P_4^{[\theta]} e^{- i \theta}$ the connection to $P_4 = i \omega_{\textbf P}$ is only directly obvious for the case that the starting value of $P_4$ is less than zero. That the same result holds for continuation from positive real $P_4$ is automatic, since no analytic structure separates the two halves of the real line.} Comparison to Eq.~\eqref{eq:CE_at_poles2} then allows us to read off the residue
\begin{align}
\mathcal M_E^{\mathcal O, (d)}(\{\textbf Q\},\textbf P',\textbf P) = {} &  (-1)^{n'} (-i)^n \int \frac{d^4\widetilde k_{1}}{(2\pi)^4}
\frac{d^4\widetilde k_{2}}{(2\pi)^4}\cdots \frac{d^4\widetilde k_{n}}{(2\pi)^4}
\,
f_M(\{\textbf Q\}, \widetilde P',\widetilde P,\widetilde k_{1},\widetilde k_{2},\ldots, \widetilde k_{n}) \nonumber \\ 
{} & \qquad \times
\,
\left[\frac{1}{{\widetilde q}^2_1-m^2_1+i\epsilon}\frac{1}{{\widetilde q}^2_2-m^2_2+i\epsilon}\cdots\frac{1}{{\widetilde q}^2_{n'}-m^2_{n'}+i\epsilon}\right]_{\widetilde P'^0 = \omega_{\textbf P'},\widetilde P^0 = \omega_{\textbf P} } \,.
\end{align}

We observe that this is equal to the Minkowski diagrammatic expression that is induced when one applies the standard Feynman rules to directly calculate a contribution to the physical matrix element. In summary, the transformation used to extract $\mathcal M_E^{\mathcal O, (d)}(\{\textbf Q\},\textbf P',\textbf P)$ generates the expression defining $\mathcal M^{\mathcal O, (d)}(\{\textbf Q\},\textbf P',\textbf P)$. We deduce that they are equal, i.e.~that Eq.~\eqref{eq:proof} is satisfied, thereby completing the proof of this section. In conclusion, we have found that large Euclidean time separation simply provides an alternative form of LSZ reduction for single-particle matrix elements.

\section{Summary\label{sec:summary}}

Quasi-distributions are a relatively new approach to determining light-front PDFs from lattice QCD. Preliminary results at a single lattice spacing have been encouraging, and 
although there are unresolved issues regarding renormalization of quasidistributions on the lattice, the approach has significant promise for first principles calculations of PDFs and GPDs. Implicit in the discussion so far has been the assumption that quasidistributions determined in Euclidean spacetime are exactly those determined in Minkowski spacetime, an assumption that has very recently been called into question.

We addressed this issue by considering the relationship of the matrix elements extracted from Euclidean correlation functions and those determined by an LSZ reduction in Minkowski spacetime. We demonstrated that the quasidistributions extracted from lattice QCD are exactly those of Minkowski spacetime. Through the example of a toy model, we illustrated how the apparent contradiction between our claim and the  perturbative analysis of \cite{Carlson:2017gpk} can be resolved. Taken together, our nonperturbative analysis and the careful examination of missing contributions in our simple model, provide strong justification for our perturbative prescription for choosing the appropriate contour in Euclidean perturbation theory (which we note has been in use in automated lattice perturbation theory calculations involving strictly local operators~\cite{Hart:2006ij}) to ensure the correct perturbative infrared behavior. {We finish by deriving an all orders analogue of the LSZ reduction formula in Euclidean spacetime, demonstrating that the matrix elements in the Euclidean and Minwkowski correlators coincide, assuming that the currents are local in time.}

\section*{Acknowledgments}

R.A.B. acknowledges support from U.S. Department
of Energy contract DE-AC05-06OR23177, under which Jefferson Science Associates,
LLC, manages and operates Jefferson Lab. The authors would like to thank C.~Carlson, M.~Freid, and K.~Orginos for pointing us to this problem, A.~Radyushkin for addressing a possible counterexample, and R.~Edwards, A. Pilloni, and J.~Qiu for useful discussions.  

\bibliography{EvsMqPDF} %%% ref.bib file

\end{document}